# Semi-Analytical Electromagnetic Transient Simulation Using Differential Transformation

Min Xiong
*Department of EECS*
University of Tennessee
Knoxville, TN, USA
mxiong3@vols.utk.edu

Rui Yao
*Argonne National Laboratory*
Lemont, IL, USA
yaorui.thu@gmail.com

Yang Liu
*Department of EECS*
University of Tennessee
Knoxville, TN, USA
yliu161@vols.utk.edu

Kai Sun
*Department of EECS*
University of Tennessee
Knoxville, TN, USA
kaisun@utk.edu

Feng Qiu
*Argonne National Laboratory*
Lemont, IL, USA
fqiu@anl.gov

*Abstract*—For electromagnetic transient (EMT) simulation of a power system, a state-space-based approach needs to solve state-space EMT equations by using numerical integration methods, e.g., the Euler method, Runge-Kutta methods, and trapezoidal-rule method, at small time steps. The simulation can be slow on a power system having multiple generators. To speed up state-space-based EMT simulations, this paper proposes a Differential Transformation based semi-analytical method that repeatedly utilizes a high-order semi-analytical solution of the EMT equations at longer time steps. The proposed semi-analytical method is tested on the detailed EMT model of a four-generator two-area system. Simulation results show the significant potential of the proposed method to accelerate EMT simulations of power systems compared with traditional numerical methods.

*Keywords—Electromagnetic transient simulation, state-space equation, differential transformation, semi-analytical solution.*

## I. Introduction

For dynamic security assessment, time-domain power system simulation tools can be categorized as electromechanical transient simulators (usually used for transient stability analysis) and electromagnetic transient (EMT) simulators. Electromechanical transient simulators use fundamental frequency based phasor models and are suitable to simulate relatively slow dynamics, e.g. transient stability of synchronous generators with a typical time step around 10 ms [1][2]. By contrast, EMT simulators use instantaneous voltage and current values and detailed three-phase models, and thus fast dynamics such as switching overvoltage, precise determination of short-circuit currents, and power quality can be studied [3].

Because the instantaneous values are used instead of phasor values, the time step required for EMT simulators is typically at μs scale, which is much smaller than that with electromechanical transient simulators. Thus, an advantage of EMT simulation is that more detailed fast dynamics can be captured. However, this also greatly increases the computation burden. Hence, EMT simulation of a large-scale power system with full EMT models is very time-consuming, if not infeasible. Meanwhile, due to the high cost of hardware-based real-time digital simulators, improving and modifying the traditional off-line EMT tools such as EMTP-RV and PSCAD to achieve real-time performance is of great interest and significance.

Approaches on fast EMT simulation can be briefly categorized as hybrid simulation, parallel computation, and shifted frequency analysis. All these approaches are based on the nodal method, in which equations of components such as inductor, capacitor, and resistor are discretized with the implicit trapezoidal-rule integration method, thus the original system can be transformed into an electric network consisting of equivalent historical source and impedance, and then the nodal equation can be easily formulated and solved at each time step. Because the nodal conductance matrix is obtained in a straightforward way to represent the network, it can adapt to system topology changes quickly. However, a small fixed time step is required to ensure accuracy because the trapezoidal-rule method is a low-order method and the time step is embedded in the conductance matrix, which adds to the computation burden.

Besides the nodal method, EMT simulation can also be done by the state-space method. After the formulation of state-space equations of a system is built, different numerical methods can be applied to get the solution and a variable time-step can be adopted to accelerate simulation. In paper [4], a combined state-space nodal method was proposed to reduce the overall computation time of the state-space method. By partitioning the system into multi sub-groups, each state-space group could formulate its state-space equations separately, thus greatly reducing the formulation time compared with a unique formulation for the whole system. The state-space equations are then discretized with trapezoidal-rule and transformed into nodal equations, which enables the simultaneous interfacing of nodal equations with state-space equations. In [5], descriptor state-space equations without elimination of dependent state variables are discretized with trapezoidal-rule and compared with the companion circuits approach. Although the overall computation time is not reduced, this method enables efficient analytical eigenvalue calculation.

The essence of EMT simulation is to solve an initial value problem of a set of differential equations that model the system and are conventionally solved by numerical solvers. Finding new, more efficient solution methods for differential equations can speed up EMT simulation. In recent years, a number of semi-analytical solution (SAS) methods different from numerical methods were proposed for fast transient stability simulation of a power system modeled by nonlinear differential and algebraic equations [6]-[10] and they

This work was supported in part by the ERC Program of the NSF and U.S. DOE under grant EEC-1041877 and in part by the Advanced Grid Modeling (AGM) program of U.S. DOE Office of Electricity under grant DE-OE0000875.

demonstrated the potential of speeding up power system time-domain simulation by using high-order semi-analytical solutions (approximate but analytical) of power system equations. However, such semi-analytical methods have not been applied to EMT simulation of a much faster times scale than the electromechanical time scale for transient stability studies. Based on the conventional state-space EMT formulation, this paper proposes a Differential Transformation (DT) based semi-analytical method to solve state-space differential equations for fast EMT simulation. DT-based semi-analytical methods have been proposed and successfully applied to electromechanical transient simulations of large-scale power systems in phasor models in the past but have not yet been validated or tested on the EMT simulation. The proposed DT method uses derived high order analytical solutions to approximate the true solutions, and therefore, much bigger time steps can be applied to speed up the simulation compared with that of traditional numerical methods for the same level of accuracy [9][10].

The rest of this paper is organized as follows. Section II introduces the DT method along with some basic transform rules. Section III presents the EMT models including synchronous generator, control components, and RLC circuits along with the DTs of some equations of those models. Section IV illustrates the steps of an EMT simulation protocol using the DT method. Section V tests the performance of the DT method-based EMT simulation on a two-area system. Conclusions and future work are discussed in section VI.

## II. INTRODUCTION OF DIFFERENTIAL TRANSFORMATION

Differential Transformation (DT) based semi-analytical methods have been developed and successfully applied to solve problems of various nonlinear dynamical systems [11] including the initial value problems and power flow problems on large-scale power system differential-algebraic equations [9][10][12]. The main advantage of the DT method is that it is very straightforward to obtain high-order approximate solutions, so-called semi-analytical solutions, in a highly efficient, recursive manner.

Considering a smooth function $f(t)$, its $k^{th}$ order DT is defined as Eq. (1) below.

$$F(k) = \frac{1}{k!}\left[\frac{d^k f(t)}{dt^k}\right]_{t=t_0} \quad (1)$$

Also, from the Taylor series, we have:

$$f(t) = \sum_{k=0}^{\infty} F(k)(t-t_0)^k \quad (2)$$

Based on the DT above, by replacing all the variables in equations with their Taylor series composed of DTs and equaling the like terms of $(t-t_0)^k$ in the equations, then linear equations of different order DTs can be established easily. Finally, solving different order DTs and plugging them back into the Taylor series gives an approximate solution, as shown in Eq. (3) below.

$$f(t) \approx \sum_{k=0}^{i} F(k)(t-t_0)^k \quad (3)$$

A set of rules for common linear and nonlinear functions were well-developed and can be used directly. Part of the rules is presented below.

Denote $f(t)$, $g(t)$ and $h(t)$ as the original continuous functions and $F(k)$, $G(k)$ and $H(k)$ as their corresponding $k^{th}$ order DTs. The following rules shown in Table I on the DTs can be obtained [9][10].

Detailed proof of these DTs can be found in [13]. More rules on nonlinear functions such as exponential function and power function can refer to [9].

## III. DTs OF POWER SYSTEM STATE-SPACE EMT MODELS

In power system EMT simulation, the differential equations can be easily transformed into DT-based equations by well-developed rules, and then it is straightforward to calculate high order DTs of state variables from their low order DTs. Finally, substituting DTs back to Eq. (2) provides an analytical approximation of state variables.

TABLE I. DIFFERENTIAL TRANSFORM FORMULAE

| Original function | Transformed function |
|---|---|
| #1 $f(t) = c$, $c$ is constant | $F(k) = c\eta(k)$, $\eta(k) = \begin{cases} 1 & k=0 \\ 0 & k \neq 0 \end{cases}$ |
| #2 $f(t) = cg(t)$ | $F(k) = cG(k)$ |
| #3 $f(t) = g(t) \pm h(t)$ | $F(k) = G(k) \pm H(k)$ |
| #4 $f(t) = g(t)h(t)$ | $F(k) = \sum_{m=0}^{k} G(k)H(k-m)$ |
| #5 $f(t) = \frac{dg(t)}{dt}$ | $F(k) = (k+1)G(k+1)$ |
| #6 $f(t) = \sin(h(t))$, $g(t) = \cos(h(t))$ | $F(k) = \sum_{m=0}^{k-1} \frac{k-m}{k} G(m)H(k-m)$, $G(k) = -\sum_{m=0}^{k-1} \frac{k-m}{k} F(m)H(k-m)$ |
| #7 $s(t) = g^2(t) + h^2(t)$, $f(t) = \sqrt{s(t)}$ | $S(k) = \sum_{m=0}^{k}(G(m)G(k-m) + H(m)H(k-m))$, $F(k) = \frac{1}{2F(0)}\left(S(k) - \sum_{m=1}^{k-1} F(m)F(k-m)\right)$ |

In this section, the applications of DT rules to equations of a detailed synchronous generator, control components, and network RLC circuits are presented in detail.

### A. Synchronous Generator Model

A detailed voltage-behind-reactance round rotor generator model is adopted as shown in (4)-(12) below [14][15].

$$\frac{d\delta}{dt} = \Delta\omega_r$$

$$\frac{d\Delta\omega_r}{dt} = \frac{\omega_0}{2H}(p_m - p_e - D\frac{\Delta\omega_r}{\omega_0})$$

$$\frac{d\lambda_{fd}}{dt} = e_{fd} - \frac{r_{fd}}{L_{lf}}(\lambda_{fd} - \lambda_{ad})$$

$$\frac{d\lambda_{1d}}{dt} = -\frac{r_{1d}}{L_{1dl}}(\lambda_{1d} - \lambda_{ad}) \quad (4)$$

$$\frac{d\lambda_{1q}}{dt} = -\frac{r_{1q}}{L_{1ql}}(\lambda_{1q} - \lambda_{aq})$$

$$\frac{d\lambda_{2q}}{dt} = -\frac{r_{2q}}{L_{2ql}}(\lambda_{2q} - \lambda_{aq})$$

$$\frac{di_{abc}}{dt} = -\frac{1}{L_{abc}^{"}}(v_{abc} - P_{ark}^{-1}v_{0dq}^{"} + R_s i_{abc} + \frac{dL_{abc}^{"}}{dt}i_{abc})$$

where $\delta$, $\omega_r$, $\Delta\omega_r$, $H$, $D$, $p_m$, and $p_e$ are rotor angle, rotor angle speed, rotor angle speed deviation, inertial constant, damping constant, mechanical power, and electrical power, respectively; $\omega_0$ is the nominal frequency of the system; $\lambda_{fd}$, $\lambda_{1d}$, $\lambda_{1q}$, and $\lambda_{2q}$ are flux linkages of filed winding, $d$-axis damper winding, $q$-axis first damper winding, $q$-axis second damper winding, respectively; $r_{fd}$, $r_{1d}$, $r_{1q}$, and $r_{2q}$ are resistances of the four windings; $L_{fdl}$, $L_{1dl}$, $L_{1ql}$, and $L_{2ql}$ are leakage inductance of the four windings; $e_{fd}$ is field voltage; $i_{abc}$ and $v_{abc}$ are three phase terminal current and voltage which interface with the grid; $R_s$ is a constant stator resistance matrix.

Also, $L''_{abc}$ is the subtransient inductance matrix shown below:

$$L''_{abc} = \begin{bmatrix} L_S(2\theta) & L_M(2\theta - 2\pi/3) & L_M(2\theta + 2\pi/3) \\ L_M(2\theta - 2\pi/3) & L_S(2\theta + 2\pi/3) & L_M(2\theta) \\ L_M(2\theta + 2\pi/3) & L_M(2\theta) & L_S(2\theta - 2\pi/3) \end{bmatrix} \quad (5)$$

where,

$$L_S(\cdot) = L_{al} + \frac{1}{3}(L_0 - L_{al} + L''_{ad} + L''_{aq}) + \frac{1}{3}(L''_{ad} - L''_{aq})\cos(\cdot)$$

$$L_M(\cdot) = \frac{1}{6}(2L_0 - 2L_{al} - L''_{ad} - L''_{aq}) + \frac{1}{3}(L''_{ad} - L''_{aq})\cos(\cdot) \quad (6)$$

$$L''_{ad} = 1/(1/L_{ad} + 1/L_{fdl} + 1/L_{1dl})$$

$$L''_{aq} = 1/(1/L_{aq} + 1/L_{1ql} + 1/L_{2ql})$$

The $d$-$q$ axis sub transient voltages $v''_d$ and $v''_q$ are:

$$v''_d = -(\frac{r_{fd}}{L^2_{fl}} + \frac{r_{1d}}{L^2_{1dl}})L''^2_{ad}i_d - (\frac{r_{fd}L''_{ad}}{L^2_{fl}}(1 - \frac{L''_{ad}}{L_{fl}}) - \frac{L''^2_{ad}}{L^2_{1dl}}\frac{r_{1d}}{L_{fl}})\lambda_{fd}$$

$$+ (\frac{L''^2_{ad}}{L^2_{fl}}\frac{r_{fd}}{L_{1dl}} - \frac{r_{1d}L''_{ad}}{L^2_{1dl}}(1 - \frac{L''_{ad}}{L_{1dl}}))\lambda_{1d} \quad (7)$$

$$- \omega_r L''_{aq}(\frac{\lambda_{1q}}{L_{1ql}} + \frac{\lambda_{2q}}{L_{2ql}}) + \frac{L''_{ad}}{L_{fl}}v_{fd}$$

$$v''_q = -(\frac{r_{1q}}{L^2_{1ql}} + \frac{r_{2q}}{L^2_{2ql}})L''^2_{aq}i_q - (\frac{r_{1q}L''_{aq}}{L^2_{1ql}}(1 - \frac{L''_{aq}}{L_{1ql}}) - \frac{L''^2_{aq}}{L^2_{2ql}}\frac{r_{2q}}{L_{1ql}})\lambda_{1q}$$

$$+ (\frac{L''^2_{aq}}{L^2_{1ql}}\frac{r_{1q}}{L_{2ql}} - \frac{r_{2q}L''_{aq}}{L^2_{2ql}}(1 - \frac{L''_{aq}}{L_{2ql}}))\lambda_{2q} \quad (8)$$

$$+ \omega_r L''_{ad}(\frac{\lambda_{fd}}{L_{fdl}} + \frac{\lambda_{1d}}{L_{1dl}})$$

The d-axis and q-axis flux linkage $\lambda_{ad}$ and $\lambda_{aq}$ are:

$$\lambda_{ad} = \frac{-i_d + \lambda_{fd}/L_{fdl} + \lambda_{1d}/L_{1dl}}{1/L_{ad} + 1/L_{fdl} + 1/L_{1dl}}$$

$$\lambda_{aq} = \frac{-i_q + \lambda_{1q}/L_{1ql} + \lambda_{2q}/L_{2ql}}{1/L_{aq} + 1/L_{1ql} + 1/L_{2ql}} \quad (9)$$

In addition, $P_{ark}$ represents the magnitude invariant Park transformation matrix which is used to transfer three phase variables to $d$-$q$ axis variables, and is shown below:

$$P_{ark} = \frac{2}{3}\begin{bmatrix} 1/2 & 1/2 & 1/2 \\ \cos(\theta) & \cos(\theta - \frac{2\pi}{3}) & \cos(\theta + \frac{2\pi}{3}) \\ -\sin(\theta) & -\sin(\theta - \frac{2\pi}{3}) & -\sin(\theta + \frac{2\pi}{3}) \end{bmatrix} \quad (10)$$

with

$$\frac{d\theta}{dt} = \omega_r \quad (11)$$

The electromagnetic power is given by equation (12) below, where $P$ is the number of poles.

$$p_e = \frac{3P}{4}\omega_r(\lambda_{ad}i_q - \lambda_{aq}i_d) \quad (12)$$

Applying the DT rules to Eqs. (4)-(12) yields the corresponding DT equations. Due to the space constraints, DT of $\lambda_{fd}$, $i_{abc}$, $v''_q$, and $p_e$ are given below, where $\psi$, $W_r$, $I$, $V$, $E_{fd}$, $P_e$ denote DTs of $\lambda$, $\omega_r$, $i$, $v$, $e_{fd}$, $p_e$, respectively. For the DTs of the Park matrix, because only constant numbers, sine functions, and cosine functions are included, they can be easily derived with rule #1 and rule #6 from Table I and are not presented in detail here. In addition, for a round rotor generator, because $L''_{ad}$ is equal to $L''_{aq}$, $L''_{abc}$ becomes a constant matrix.

$$(k+1)\psi_{fd}(k+1) = E_{fd}(k) - \frac{r_{fd}}{L_{lf}}(\psi_{fd}(k) - \psi_{ad}(k)) \quad (13)$$

$$(k+1)I_{abc}(k+1) = \frac{-1}{L''_{abc}}(V_{abc}(k) + R_s I_{abc}(k)) - \sum_{m=0}^{k}(P^{-1}_{ark}(m)V''_{0dq}(k-m))) \quad (14)$$

$$V''_q(k+1) = -(\frac{r_{1q}}{L^2_{1ql}} + \frac{r_{2q}}{L^2_{2ql}})L''^2_{aq}I_q(k+1)$$

$$- (\frac{r_{1q}L''_{aq}}{L^2_{1ql}}(1 - \frac{L''_{aq}}{L_{1ql}}) - \frac{L''^2_{aq}}{L^2_{2ql}}\frac{r_{2q}}{L_{1ql}})\psi_{1q}(k+1) \quad (15)$$

$$+ (\frac{L''^2_{aq}}{L^2_{1ql}}\frac{r_{1q}}{L_{2ql}} - \frac{r_{2q}L''_{aq}}{L^2_{2ql}}(1 - \frac{L''_{aq}}{L_{2ql}}))\psi_{2q}(k+1)$$

$$+ L''_{ad}\sum_{m=0}^{k+1}(W_r(m)(\frac{\psi_{fd}(k+1-m)}{L_{fdl}} + \frac{\psi_{1d}(k+1-m)}{L_{1dl}}))$$

$$P_e(k+1) = \frac{3P}{4}\sum_{m=0}^{k+1}((W_r(m)\sum_{n=0}^{k+1-m} (\psi_{ad}(n)I_q(k+1-m-n) - \psi_{aq}(n)I_d(k+1-m-n)) \quad (16)$$

### B. Turbine-Governing Model

The adopted TGOV1 model is shown in Eq. (17) [16].

$$V_{\min} \leq p_1 \leq V_{\max}$$

$$\frac{dp_1}{dt} = \frac{1}{T_1}(\frac{1}{R}(p_{ref} - \Delta\omega_r) - p_1)$$

$$\frac{dp_2}{dt} = \frac{1}{T_3}(T_2\frac{dp_1}{dt} + p_1 - p_2) \quad (17)$$

$$p_m = p_2 - D_t\Delta\omega_r$$

where $p_m$ is the mechanical power; $p_1$ and $p_2$ are intermediate state variables; $p_{ref}$ is the reference power, and others are control parameters.

The corresponding DT of Eq. (17) is given below:

$$\begin{cases} P_1(k+1)=0, \text{ if } p_1=V_{max} \text{ and } \frac{dp_1}{dt}>0, \text{ or } p_1=V_{min} \text{ and } \frac{dp_1}{dt}<0 \\ (k+1)P_1(k+1)=\frac{1}{T_1}(\frac{1}{R}(\eta(k)p_{ref}-\Delta W_r(k))-P_1(k)), \text{ otherwise} \end{cases}$$

$$(k+1)P_2(k+1)=\frac{1}{T_3}((k+1)T_2P_1(k+1)+P_1(k)-P_2(k))$$

$$P_m(k+1)=P_2(k+1)-D_t\Delta W_r(k+1)$$

(18)

*C. Exciter Model*

The SEXS Exciter is used and described below [17]:

$$E_{min}\le e_{fd}\le E_{max}$$
$$\frac{de_{fd}}{dt}=\frac{1}{T_E}(k_E v_3 - e_{fd})$$
$$\frac{dv_3}{dt}=\frac{1}{T_B}(T_A\frac{dv_2}{dt}+v_2-v_3)$$
$$v_2 = v_{ref} - v_t$$
$$v_t = (v_d^2 + v_q^2)^{1/2}$$

(19)

where $v_{ref}$ is voltage regulator reference; $v_t$ is terminal voltage; $v_2$ and $v_3$ are intermediate variables; $K_E$, $T_E$, $T_A$, $T_B$ are exciter control parameters.

The DT of (19) is given in Eq. (20) below.

$$\begin{cases} E_{fd}(k+1)=0, \text{ if } e_{fd}=E_{max} \text{ and } \frac{de_{fd}}{dt}>0, \text{ or } e_{fd}=E_{min} \text{ and } \frac{de_{fd}}{dt}<0 \\ (k+1)E_{fd}(k+1)=\frac{1}{T_E}(k_E V_3(k)-E_{fd}(k)), \text{ otherwise} \end{cases}$$

$$(k+1)V_3(k+1)=\frac{1}{T_B}((k+1)T_A V_2(k+1)+V_2(k)-V_3(k))$$

$$V_2(k+1)=\eta(k+1)v_{ref}-V_t(k+1)$$

$$V_t(k+1)=\frac{1}{2V(0)}\left(U(k+1)-\sum_{m=1}^{k}V_t(m)V_t(k-m)\right)$$

$$U(k+1)=\sum_{m=0}^{k+1}\left(V_d(m)V_d(k+1-m)+V_q(m)V_q(k+1-m)\right)$$

(20)

*E. Network RLC(Resistor Inductor Capacitor) circuits Model*

For components including transmission lines represented by a PI section, transformers, constant impedance loads, and fixed shunts, they can all be regarded as RLC circuits and be divided into grounding capacitance circuits and serial resistance-inductance circuits generally.

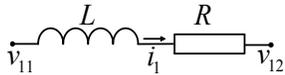

Fig. 1. A serial resistor-inductor circuit

For a resistor-inductor circuit shown in Fig. 1, the corresponding equation is:

$$\frac{di_1}{dt}=L^{-1}(v_{11}-v_{12}-Ri_1)$$

(21)

The DT of Eq. (21) is given below:

$$(k+1)I_1(k+1)=L^{-1}(V_{11}(k)-V_{12}(k)-RI_1(k))$$

(22)

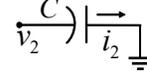

Fig. 2. A grounding capacitance circuit

For a grounding capacitance shown in Fig. 2, its equation is:

$$\frac{dv_2}{dt}=C^{-1}i_2$$

(23)

The DT of (23) is given below:

$$(k+1)V_2(k+1)=C^{-1}I_2(k)$$

(24)

Note that in three-phase AC systems, currents and voltages are three-phase values, and hence $R$, $L$, $C$ parameters are 3×3 matrices.

IV. PROPOSED DT-BASED EMT SIMULATION PROTOCOL

Based on the DT equations derived in the last section, the simulation steps are generalized below:

1) Separate all state variables into three groups:

   $x_1(t)=[\delta, \Delta\omega_r, \lambda_{fd}, \lambda_{1d}, \lambda_{1q}, \lambda_{2q}, i_{abc}, \theta, p_1, p_2, e_{fd}, v_3]$

   $x_2(t)=[i_{net}, v_{net}]$

   $x_3(t)=[i_d, i_q, v_d, v_q, v_t, \lambda_{ad}, \lambda_{aq}, p_e, v_3, v_t, v_d'', v_q'']$

   where $x_1$ includes all generator state variables; $i_{net}$ includes all network inductor currents; $v_{net}$ includes all network capacitance voltages; $x_3$ includes all terms which can be calculated by algebraic equations. Also, because only balanced faults are considered in this paper, both $i_0$ and $v_0$ are equal to 0 and are not chosen as state variables.

2) Obtain initial values of all state variables at $t=t_0$, and denote them as $X_1(0)$, $X_2(0)$, $X_3(0)$;

3) Calculate $X_1(1)$ with corresponding DT of Eq.(4); calculate $X_2(1)$ with Eq.(22) and (24);

4) Calculate $X_3(1)$ by the corresponding DT equations of $x_3$ variables;

5) Repeat step 2)-4) to calculate $X_1(2)$, $X_2(2)$, $X_3(2)$ using $X_1(0-1)$, $X_2(0-1)$, $X_3(0-1)$;

6) Continue the protocol of step 2)-4) to calculate DTs up to the $k^{th}$ order, which are $X_1(k)$, $X_2(k)$, $X_3(k)$;

7) Calculate the approximate value of $x_1(\Delta t)$, $x_2(\Delta t)$, $x_3(\Delta t)$ using Eq. (3), where $\Delta t$ is the simulation time step;

8) Update $t_0=t_0+\Delta t$, and use $x_1(\Delta t)$, $x_2(\Delta t)$, $x_3(\Delta t)$ as new initial values, then repeat step 2)-7) to calculate $x_1(2\Delta t)$, $x_2(2\Delta t)$, $x_3(2\Delta t)$;

9) Continue the multi-stage protocol of step 8) to calculate $x_1(t)$, $x_2(t)$, $x_3(t)$ at further time steps till finish the simulation.

Generally, a higher-order DT, e.g., the value of $k$ in Eq. (3), enables a larger time step but will increase the computation burden at each step.

## V. CASE STUDIES ON A TWO-AREA SYSTEM

This section tests the performance of the DT method-based EMT simulation protocol on a two-area system as shown in Fig. 3 [18].

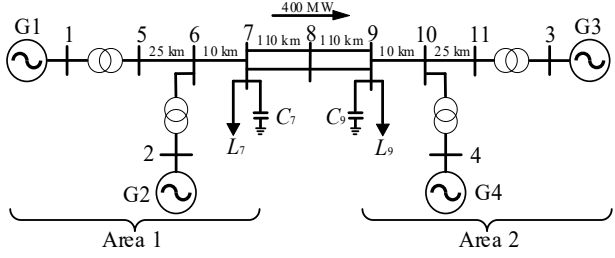

Fig. 3. Two-area system

In the simulation, the system is in a steady state at first, and at $t=1$ s, bus 7 is grounded and lasted for 5 cycles. The fault is then cleared without tripping any lines. In this test, the benchmark simulation result is given by the $4^{th}$ order Runge-Kutta (RK4) method using a tiny time step of 1 μs.

After that, the 2 seconds post fault simulation result provided by the $20^{th}$-order DT method using a time step of 100 μs is compared with the benchmark in Figures. 4-7.

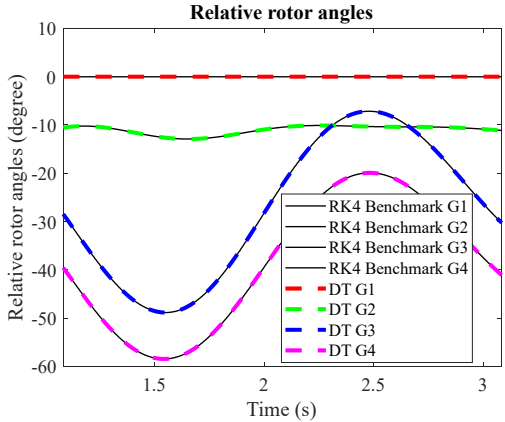

Fig.4. Simulation result of relative rotor angle

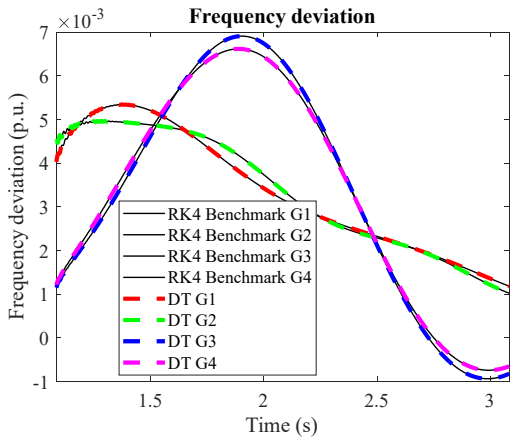

Fig.5. Frequency deviation of generators

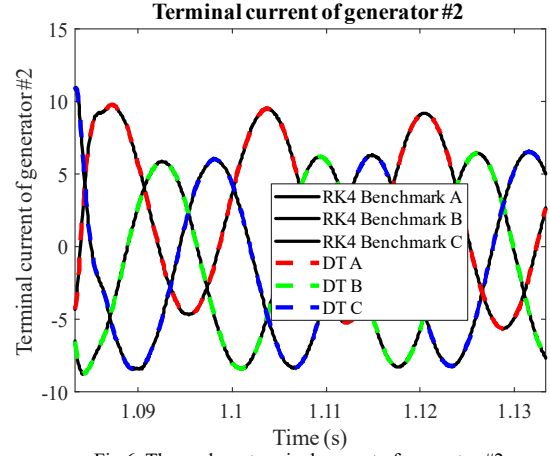

Fig.6. Three-phase terminal current of generator #2

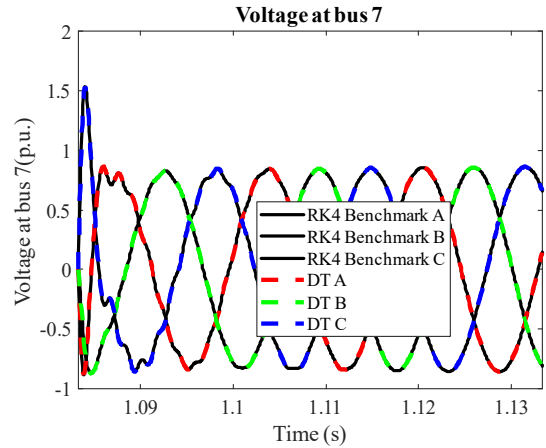

Fig.7. Three-phase voltage at bus #7

From Figs. 4-7, results given by the DT method accurately match the benchmark result. Also, convergence and accuracy are ensured when the time step is not too large.

Then, under different error tolerances, the computation efficiency of the DT method using different orders on the post fault simulation lasts for 2 seconds are presented in Table II. Finally, the computation cost of the $30^{th}$-order DT is compared with that of the RK4 method, modified Euler method (ME), and trapezoidal-rule method under different error tolerances. Here, an error tolerance is provided by the maximum absolute error of all state variables over the whole simulation time interval compared with the benchmark results. The comparison results are given in Table III.

TABLE II. TIME STEPS AND COSTS OF DIFFERENT ORDER DT METHODS

| Error (p.u.) | DT ($10^{th}$ order) | | DT ($20^{th}$ order) | | DT ($30^{th}$ order) | | DT ($40^{th}$ order) | |
|---|---|---|---|---|---|---|---|---|
| | Time step (μs) | Time cost (s) | Time step (μs) | Time cost (s) | Time step (μs) | Time cost (s) | Time step (μs) | Time cost (s) |
| $10^{-2}$ | 72 | 81 | 200 | 58 | 330 | 53 | 412 | 58 |
| $10^{-3}$ | 58 | 100 | 180 | 64 | 306 | 57 | 382 | 62 |
| $10^{-4}$ | 46 | 1162 | 158 | 73 | 284 | 62 | 365 | 68 |

TABLE III. COMPARING TIME STEPS AND COSTS OF DIFFERENT METHODS

| Error (p.u.) | ME | | RK 4 | | DT (30$^{th}$ order) | | Trapezoidal-rule | |
|---|---|---|---|---|---|---|---|---|
| | Time step (μs) | Time cost (s) | Time step (μs) | Time cost (s) | Time step (μs) | Time cost (s) | Time step (μs) | Time cost (s) |
| $10^{-2}$ | 1.0 | 1045 | 10 | 200 | 330 | 53 | 400 | 92 |
| $10^{-3}$ | 0.5 | 2100 | 5 | 360 | 306 | 57 | 400 | 230 |
| $10^{-4}$ | 0.5 | 4215 | 3 | 550 | 284 | 62 | 400 | 600 |

From Table II, with the increase of the order of DT, the time step keep increasing, and the computation cost first decreases and then increases. This validates that high order DT could enlarge the time step, but when the order is too high, the burden of computing high order terms would decrease the overall computation efficiency.

From Table III, the time step of the 30$^{th}$ order DT method is much bigger than that of the ME and RK4 methods, thus the simulation time cost of the DT method is much smaller especially when the simulation desires smaller error tolerance. Also, the time step of the trapezoidal-rule method (utilized by ode23t in MATLAB with the setting of error limit) does not affect its computation cost because it is an implicit method, and iterations are needed to converge to an accurate result, and more iterations are required when the time step is big. Thus, the overall computation remains the same and only the results under a 400 μs time step are given in the table. Specifically, the computation speed of 30$^{th}$ order DT is 4 times or more faster than other methods when the error tolerance is $10^{-3}$, and 8.8 times or more faster when the error tolerance is $10^{-4}$. Thus, the simulation results validate the high computation efficiency and accuracy of the proposed DT-based approach.

## VI. CONCLUSION

This paper proposes a differential transformation (DT)-based semi-analytical method for EMT simulation in the state space. Based on well-developed DT transform rules, the DTs of EMT state-space equations considering various common system components are derived. Because high-order series terms are used in the DT method, the time step can be significantly enlarged, and thus the time cost can be reduced. The simulation results on the two-area system demonstrate that the DT-based approach has achieved a large time step and reduced the time cost significantly for a small error tolerance at the level of $10^{-4}$. The proposed method can also be utilized in the simulation of renewable energy resources, distributed transmission lines, unbalanced faults, and will be developed in our future work.